\documentclass[%
 reprint,
 amsmath,amssymb,
 aps,
prl
]{revtex4-2}

\usepackage{graphicx}
\usepackage{float}
\usepackage{verbatim}
\usepackage{dcolumn}
\usepackage{physics}
\usepackage{bm}
\usepackage{lipsum}

\begin{document}

\preprint{APS/123-QED}

\title{Time-resolved quantum correlations in electronic noise}

\author{Jean-Olivier Simoneau}
\email{now at Nord Quantique, nordquantique.ca}
\author{S\'ebastien Jezouin}
\email{now at Alice $\&$ Bob,  alice-bob.com}
\author{Christian Lupien}
\author{Bertrand Reulet}%

\affiliation{Département de physique, Institut quantique, Université de Sherbrooke, Sherbrooke, Québec, Canada J1K 2R1
}%

\date{\today}

\begin{abstract}

The statistics of quantum transport in nanostructures can be tailored by a time-dependent bias voltage $V(t)$. We demonstrate experimentally how correlations of current fluctuations at two different times $t$ and $t+\tau$ depend on the shape of $V(t)$ via the phase accumulated by the electronic wavefunctions between $t$ and $t+\tau$. For this we measure the current-current correlation of the shot noise of an ac+dc biased tunnel junction using a 10 GHz bandwidth, time-resolved detection. Our result allows to explore correlations within a single excitation period. It demonstrates the counterpart of the ac Josephson effect in superconducting junctions, to a normal, non-superconducting mesoscopic device. 

\end{abstract}

\maketitle

Probing electron transport in coherent conductors beyond the average current is essential to capture how quantum correlations affect the statistics of electrical current. The effect of a time-dependent voltage $V(t)$ on the current-current correlator has attracted a long lasting, both experimental and theoretical interest. The current noise spectral density at frequency $f=\omega/2\pi$, i.e. $S(\omega)=\ev{I(-\omega)I(\omega)}$ in the presence of an ac excitation of various shape (sinusoidal\cite{Schoelkopf1998}, shifted sinusoidal \cite{Gasse2014}, bi-harmonic\cite{Gabelli_shaping}, Lorentzian\cite{levitons}) at another frequency $f_0=\Omega/2\pi$ has been observed in various quantum systems (tunnel junction, diffusive wire, quantum Hall edge states, etc.) and interpreted in terms of photo-assisted noise \cite{Schoelkopf1998,Gasse2014}, spectroscopy of the distribution function \cite{Gabelli_shaping, levitons}, as well as quantum tomography of electronic signals \cite{Jullien2014, Bisognin2019}. The same ac excitation has been shown to induce correlations between currents at different frequencies, captured by the noise susceptibilities $\beta_n(\omega)= \ev{I(\omega)I(n\Omega-\omega)}$ with $n$ integer\cite{Gabelli2008,Farley2023}, leading to the generation of squeezed vacuum in the microwave domain\cite{Gasse2013}. All these experiments can be accounted for by a single, simple formula which is enlightening in time-domain\cite{Altshuler}. For a conductor with low transmission like a tunnel junction, it reads:
\begin{equation}
\ev{I(t)I(t+\tau)}=S_{\textrm{eq}}(\tau)\cos [{\Phi(t,t+\tau)}]   
\label{Eq_main}
\end{equation}
with 
\begin{equation}
\Phi(t,t+\tau)=\frac e\hbar\int_t^{t+\tau} V(t')dt' 
\label{eq_Phi}
\end{equation}
the phase accumulated by electronic wave functions between $t$ and $t+\tau$ due to the applied voltage, and $S_{\textrm{eq}}(\tau)$ the current-current correlator at equilibrium, independent of $t$ (for a general conductor, one has to multiply the r.h.s. of Eq.(\ref{Eq_main}) by the Fano factor $F$, and add the voltage-independent contribution $(1-F)S_{\textrm{eq}}(\tau)$, to recover the equilibrium noise when $V=0$). These formulas are direct consequences of the voltage across the sample acting on the phase of the electronic wavefunctions\cite{TienGordon}. They bear strong similarities with Josephson equations for the supercurrent $I_s(t)$ in a voltage-biased superconducting tunnel junction: $I_s(t)=I_c\sin\Phi(t)$ with $\Phi(t)$ given by Eq.(\ref{eq_Phi}) where $e$ has to be replaced by $2e$, the charge of a Cooper pair, and the integral spans from $-\infty$ to $t$. In a superconductor all the electrons have the same phase and the action of the voltage on that phase is visible in the average current. In a normal metal all electrons have different phases but the voltage acts equally on all of them. As a consequence, the average current follows Ohm's law, but according to Eq.(\ref{Eq_main}) the current-current correlator reveals the direct coupling between the applied voltage and the phase of the electronic wavefunctions  (in the particular case of a ballistic system, a single voltage pulse has been predicted to generate oscillations of the current analoguous to the ac Josephson effect \cite{Waintal_JJac}). While causality implies that the current-current correlator cannot depend on $V$ after $t+\tau$, nothing prevents it to depend on the past of the voltage before $t$. Eq.(\ref{Eq_main}) claims that it does not: only the voltage between $t$ and $t+\tau$ matters. In contrast, the supercurrent at any time in the Josephson junction depends on the voltage at all times in the past.

Usual experiments, by averaging over time, have observed \emph{consequences} of Eq.(\ref{Eq_main}) integrated over $t$. In particular, the usual noise spectral density $S(\omega)=\ev{I(\omega)I(-\omega)}$ corresponds to the Fourier transform with respect to $\tau$ of $\ev{I(t)I(t+\tau)}$ averaged over the time $t$. From the frequency dependence of $S(\omega)$, the $\tau$-dependence of the current-current correlator can be reconstructed, leading to the observation of Pauli-Heisenberg oscillations for a time-independent bias voltage\cite{Karl}. But the dependence on $t$ cannot be recovered from these experiments, and as far as we know Eq.(\ref{Eq_main}) has never been experimentally validated for a time-dependent voltage.
Here we provide the time-domain measurement of current-current correlators synchronously with an ac sinusoidal excitation at frequency $f_0=4$GHz. This allows us to explore correlations in current fluctuations inside a single excitation period, to define phase-dependent noise and deduce noise susceptibilities and time / frequency Wigner functions, all in agreement with predictions from Eq.(\ref{Eq_main}).

\emph{Phase-dependent noise.} %
In the following we show an experiment that aims at demonstrating the validity of Eq.(\ref{Eq_main}). In order to measure the average current-current correlator, one must experimentally use a periodic excitation and take the average over many repetitions of the same excitation. We have chosen to work with a sine-wave excitation on which a dc voltage is superimposed, which is enough for our goal: $V(t)=V_{\textrm{dc}}+V_{\textrm{ac}}\cos\Omega t$. The current-current correlator $\ev{I(t)I(t+\tau)}$ is also periodic in $t$, so the initial time $t$ plays the role of a phase $\phi=\Omega t$. We define the phase-dependent noise $S_\phi(\tau)=\ev{I(\phi/\Omega)I(\phi/\Omega+\tau)}$, which can be expanded in terms of a Fourier series:
\begin{equation}
    S_\phi(\tau)=\sum_n\beta_n(\tau)e^{i n\phi}
\end{equation}
The real functions $\textrm{Re}[\beta_n(\tau)]$ and $\textrm{Im}[\beta_n(\tau)]$ represent the nth harmonic of the in-phase and in-quadrature response of the noise to the ac excitation. We will also consider the Fourier transform of the phase-dependent noise, $S_\phi(\omega)$, which represents a phase-dependent spectrum. The noise spectral density $S(\omega)$ usually considered in the literature corresponds to the average of $S_\phi(\omega)$ over $\phi$. $S_\phi$ is $2\pi$-periodic in $\phi$, and its Fourier coefficients are given by: $\beta_n(\omega)=\ev{I(\omega)I(n\Omega-\omega)}$ i.e., they are the noise susceptibilities, which measure the correlations between two Fourier components of the current at frequencies separated by $n\Omega$ \cite{Gabelli2008,Gabelli_SPIE}. And $\beta_0 (\omega)=S(\omega)$. 

\begin{figure}
    \centering
    \includegraphics[width=\columnwidth]{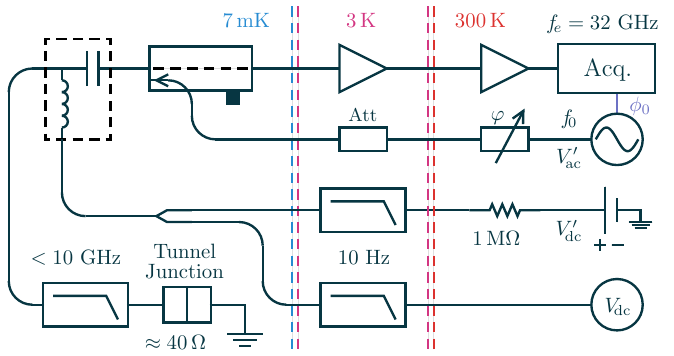}
    \caption{Schematics of the experimental setup. The $\varphi$ symbol represents a mechanical variable delay, the Att. symbol the attenuation along the excitation line, which is distributed at different temperatures.}
    \label{Fig_schema}
\end{figure}

\emph{Wigner distribution function.} %
While the definition of the phase-dependent noise spectra presented above is quite natural and makes a clear link with the noise susceptibilities, it suffers from being a complex quantity, unlike $S(\omega)$, since in general $\ev{I(t)I(t+\tau)}$ is not an even function of $\tau$. The Wigner distribution function $W(t,\omega)$ of the time-dependent current $I(t)$, defined as the Fourier transform with respect to $\tau$ of $\ev{I(t-\tau/2)I(t+\tau/2)}$ is also a real quantity, closely related to $S_\phi(\omega)$. It is given by:
\begin{equation}
    W(t,\omega)=\sum_nM_n(\omega)e^{i n\Omega t}
    \label{eq:Wigner}
\end{equation}
with $M_n(\omega)=\beta_n(\omega+n\Omega/2)$. $W(t,\omega)$ is indeed real since $\beta_{n}(\omega)=\beta_n(n\Omega-\omega)=\beta_{-n}(-\omega)^*$. We discuss later its link with the electronic Wigner function, another quantity that has been shown to be a key concept to perform tomography of quantum signals\cite{Ferraro2013, Jullien2014, Bisognin2019, Roussel2021}.


\begin{figure}
    \centering
    \includegraphics[width=\columnwidth]{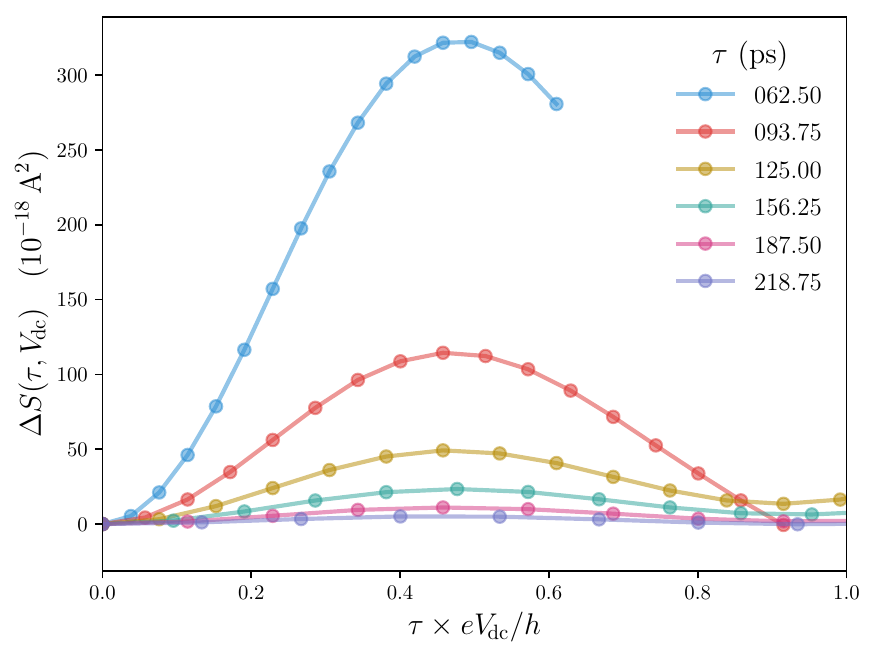}
    \caption{Voltage excess noise $\Delta S(\tau,V_\textrm{dc})$ as a function of $eV_\textrm{dc}\tau/h=\Phi_\textrm{dc}/(2\pi)$ for various times $\tau$ in the absence of an ac excitation.}
    \label{Fig_S_excess_dc}
\end{figure}

\emph{Experiment.} %
The experiment, presented in Fig.\ref{Fig_schema}, is conceptually very simple: the sample, a tunnel junction of dc resistance $40\Omega$, is ac+dc biased through a microwave coupler and bias tee; the noise it generates is amplified, digitized and recorded. From the time traces we calculate the correlators $S_\phi(\tau)$. The ac excitation is at frequency $f_0=4$GHz and the digitization takes place at a pace of 32GS/s, so we have access to 8 points per cycle, i.e. 8 values of $\phi$ (or a time resolution of 31.25ps). For each $\phi$ the current-current correlator is computed on the fly for 128 values of $\tau$, i.e. from -2 to +2ns, which is much greater than the decay time of the correlations \cite{JO_code}. It is then averaged over many repetitions. Since the data averaging takes days, it is crucial to ensure that the digitization clock and the excitation of the sample remain phase-locked over very long periods. To achieve this we regularly measure the phase of the 4GHz excitation that is reflected on the sample and digitized together with the sample noise. The phase drift is calculated then corrected by a motorized mechanical phase shifter on the excitation line. This ensures a phase variation of less than one degree over weeks. A special care has to be taken in the calibration process. As usual in noise measurements, the power gain and noise temperature of the setup are obtained by applying a large dc voltage on the sample and measuring noise vs. voltage\cite{Spietz2003}. In this limit the noise of the junction is simply the classical shot noise $eV_\textrm{dc}/R$. But time domain measurements are affected not only by the gain or attenuation of the setup, but also by its phase response. Indeed, noise susceptibilities involve the complex gain at different frequencies, and the frequency-dependent phase response of the setup matters. To calibrate out that effect we again consider the classical limit of shot noise. In this limit the noise of the junction follows instantaneously the applied voltage, i.e. $\ev{I(t)I(t+\tau)}=eV(t)\delta(\tau)/R$. Thus the noise susceptibilities are real and independent of frequency: $\beta_n(\omega)=e/R$. Measuring the noise susceptibilities in this regime provides a calibration of the phase response of the experimental setup. The averaged current-current correlator is deconvolved by the response function of the setup to provide the calibrated experimental data shown in the following.


\emph{Results: dc bias only.} %
In the absence of ac bias, $V_\textrm{ac}=0$, our experiment allows for a wide bandwidth measurement of the noise spectral density $S(\omega)$ vs. frequency in the presence of a dc voltage, as in \cite{Karl}. From these data we can compute the so-called voltage excess noise $\Delta S(\tau,V_\textrm{dc})=S(\tau,V_\textrm{dc})-S(\tau,V_\textrm{dc}=0)$, with $S(\tau,V_\textrm{dc}=0)=S_\textrm{eq}(\tau)$. This quantity, once put in time domain, has been shown theoretically a long time ago to oscillate as\cite{theory_excess}:
\begin{equation}
    \Delta S(\tau,V_\textrm{dc})=-S_\textrm{eq}(\tau)\sin^2\Phi_\textrm{dc}
\end{equation}
which is nothing but a special case of Eq.(\ref{Eq_main}) for a constant bias voltage. $\Phi_\textrm{dc}$ is the phase generated by the dc voltage, $\Phi_\textrm{dc}=eV_\textrm{dc}\tau/\hbar$. We provide the experimental validation of this formula in Fig.\ref{Fig_S_excess_dc}. These data have been obtained by sweeping the dc voltage bias and measure the current-current correlations vs. $\tau$ for each voltage. They make it clear that the voltage enters via the flux $\Phi_\textrm{dc}$. From these data, taking $\Phi_\textrm{dc}=\pi/2$ provides a measurement of the equilibrium noise in time domain. 
At zero temperature, $S_\textrm{eq}(\tau)=-\hbar/\tau^2$. The divergence at short time relates to the noise spectral density behaving as $G\hbar\omega$ at high frequency. It corresponds to the zero-point motion of electrons. $S_\textrm{eq}(\tau)$ is negative at short time as a result of the Pauli exclusion principle.

\begin{figure}
    \centering
    \includegraphics[width=\columnwidth]{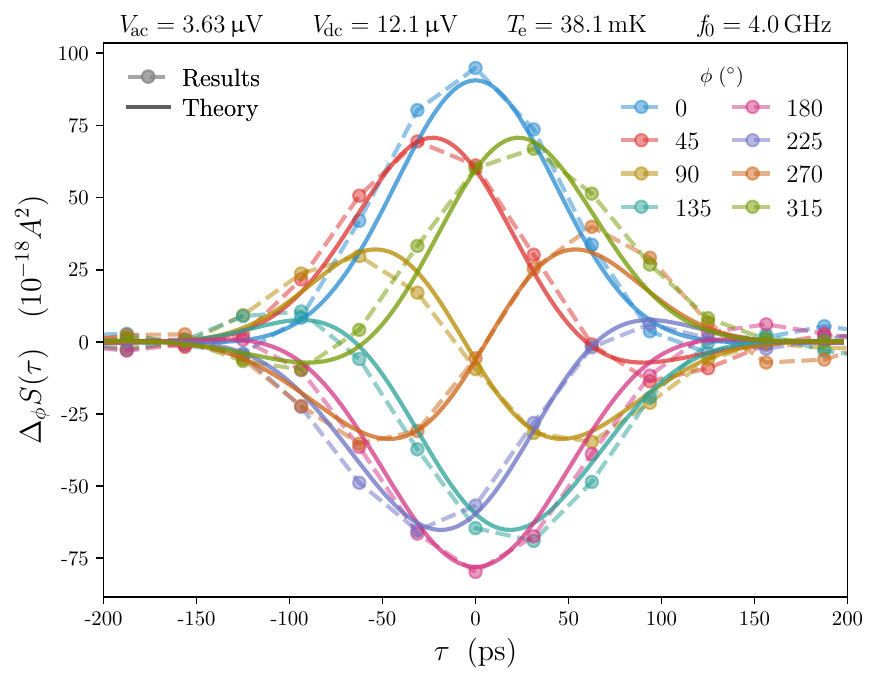}
    \caption{Phase-dependent part of the time-domain noise $S_\phi(\tau)$ vs. time $\tau$ for $V_\textrm{dc}=12.1\mu$V and $V_\textrm{ac}=3.63\mu$V.  Symbols linked by dashed lines are experimental data, solid lines are theoretical prediction of Eq.(\ref{Eq_main}) for $T=38$mK.}
    \label{Fig_Delta_phi_S}
\end{figure}

\emph{Phase resolved noise.} %
To focus on the phase-dependence of the noise we remove the phase-independent part $\beta_0=S$ to define, both in time- and frequency domains, $\Delta_\phi S=S_\phi-S$, which is shown in Fig. \ref{Fig_Delta_phi_S} as a function of $\tau$ for different phases $\phi$.  Agreement between experiment (symbols and dashed lines) and theoretical predictions of Eq.(\ref{Eq_main}) (solid lines) is very good. %
In order to get a deeper insight into the observed behaviour of $\Delta_\phi S(\tau)$, we calculate it for a small ac excitation and find:
\begin{equation}
    \Delta_\phi S(\tau)\simeq S_\textrm{eq}(\tau)\Phi_\textrm{ac}(\phi,\tau)\sin\Phi_\textrm{dc}(\tau)
    \label{eq:theory_ac}
\end{equation}
It is a combination of an oscillation vs. the dc flux, an oscillation due to the ac excitation, since $\Phi_\textrm{ac}=z\sin(\Omega\tau/2)\cos(\phi+\Omega\tau/2)$, and a decay. Here $z=eV_\textrm{ac}/(\hbar\Omega)$.

The theory in Fig. \ref{Fig_Delta_phi_S} is very close to Eq.(\ref{eq:theory_ac}) since $|\Phi_\textrm{ac}|\leq z=0.24\ll2\pi$. The curves show various symmetries: they are even functions of $\tau$ for $\phi=0$ and $\pi$, and odd for $\phi=\pi/2$ and $3\pi/2$. This comes from the symmetries in the excitation voltage: $\cos(\phi+\Omega\tau)=\cos(\phi-\Omega\tau)$ for $\phi=0$ or $\pi$, and $\cos(\phi+\Omega\tau)=-\cos(\phi-\Omega\tau)$ for $\phi=\pi/2$ or $3\pi/2$. There is no particular symmetry for $\phi=\pi/4$. The curves have crossing points. This comes from the total phase $\Phi=\Phi_\textrm{dc}+\Phi_\textrm{ac}$ accumulated by the voltage being the same for these curves, for example $\phi=0$ and $\phi=-\pi/4$ correspond to the same phase for $\Omega\tau=\pi/4$. However in the first case the ac voltage decays from its maximum value $V_\textrm{ac}$ to $V_\textrm{ac}/\sqrt{2}$ whereas in the second it grows from $V_\textrm{ac}/\sqrt{2}$ and keeps increasing to reach $V_\textrm{ac}$. In both cases the past of the time-dependent voltage is totally different. Yet $S_\phi$ is the same, in agreement with the fact that only the voltage between $t$ and $t+\tau$ matters via the accumulated phase.

\begin{figure}
    \centering
    \includegraphics[width=\columnwidth]{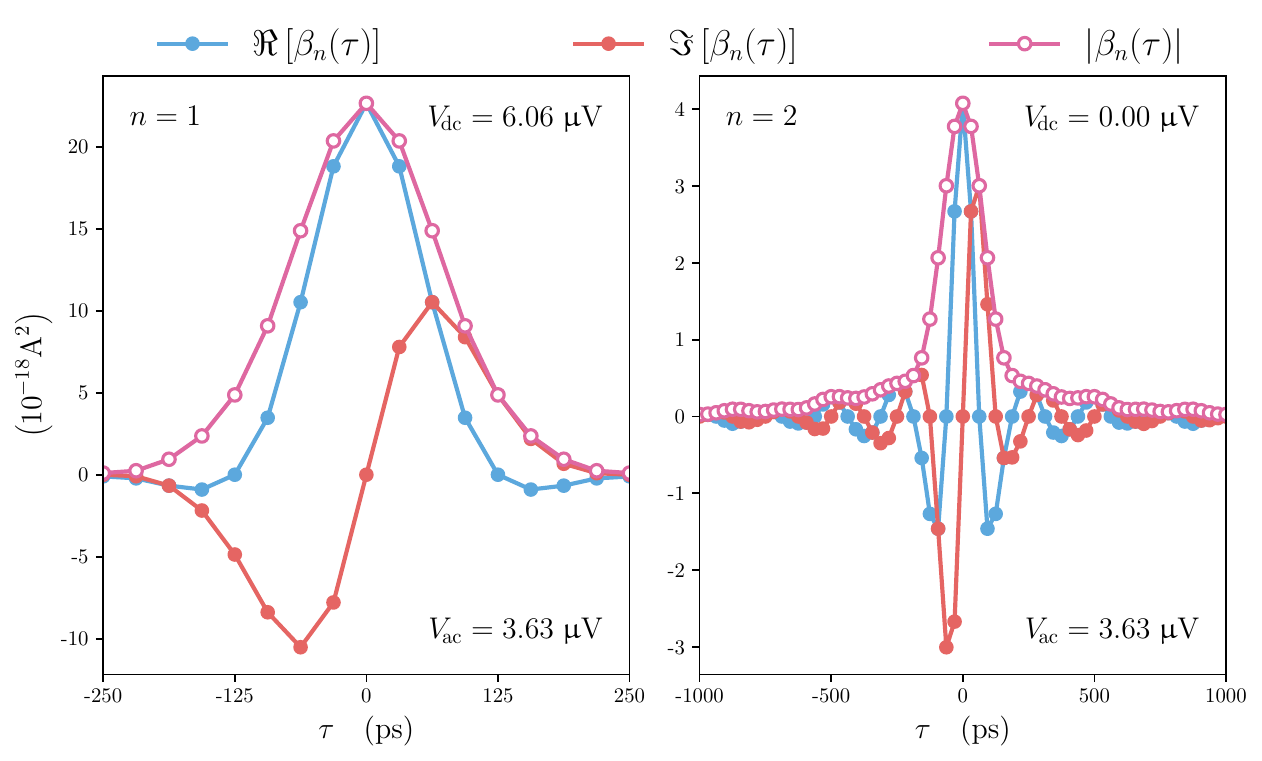}
    \caption{Measured real part, imaginary part and modulus of the time-domain harmonic responses $\beta_n(\tau)$ as a function of the time $\tau$ for $n=1$ and    $V_\textrm{dc}=6.06\mu$V (left), and for $n=2$ and $V_\textrm{dc}=0$ (right).}
    \label{Fig_XY}
\end{figure}

In Fig. \ref{Fig_XY} we show the real part, imaginary part and modulus of $\beta_n(\tau)$, i.e. the noise susceptibility in time domain, for $n=1$ at $V_\textrm{dc}\neq0$ (left) and $n=2$ at $V_\textrm{dc}=0$ (right). At equilibrium, $V(t)$ oscillates between $\pm V_\textrm{ac}$ at frequency $\Omega$. One expects the noise to respond at twice that frequency, i.e. $\beta_1=0$, thus we show the case $n=2$ at zero dc bias (we indeed observe $\beta_1\simeq0$). Both real and imaginary parts of $\beta_n(\tau)$ show a fast oscillation modulated by a slow decay. The modulus $|\beta_n(\tau)|$ decays continuously at long times without oscillating. In order to understand this behaviour we introduce the coefficients $M_n(\tau)$ of the Fourier series decomposition of the Wigner function $W(t,\tau)$:
\begin{equation}
    W(t,\tau)=\sum_n M_n(\tau)e^{i n\Omega t}
\end{equation}
$M_n(\tau)$ are the Fourier transform of $M_n(\omega)$, see Eq.(\ref{eq:Wigner}). The definitions of $S_\phi$ and $W$ imply: $\beta_n(\tau)=M_n(\tau)\exp(in\Omega\tau/2)$. From Eq.(\ref{Eq_main}), it is easy to show that for an excitation that has a symmetry point in time, i.e. $V(-t)=V(t)$, $M_n(\tau)$ is real. This implies that $\beta_n(\tau)$ is purely real for $\tau=0$ and alternates from being purely real to purely imaginary every $\tau=\pi/\Omega$, while $|M_n(\tau)|=|\beta_n(\tau)|$ acts as an envelope for the oscillation at frequency $n\Omega/2$ of $\beta_n(\tau)$. This is precisely what is observed in Fig. \ref{Fig_XY} with $\pi/\Omega=125$ps, i.e. half the period of $V(t)$. Oscillations occur indeed twice as fast for $n=2$ (right) than for $n=1$ (left).

Figs. \ref{Fig_Delta_phi_S} and \ref{Fig_XY} report time-domain, direct measurements of the Wigner distribution function of the shot noise generated by the tunnel junction. This is the Wigner function of the detected \emph{signal}. A few experiments have achieved the measurement of the \emph{electronic} Wigner function $W_e(t,\omega)$ \cite{Jullien2014, Bisognin2019} in quantum Hall edge channels, by measuring the shift in zero frequency noise caused by the addition of a small ac sinusoidal signal at frequency $n\Omega$, as suggested in \cite{Grenier2011}. 
The extra ac excitation causes a change in $W(t,\tau)$, with a modulation in $\tau$ at frequency $n\Omega/2$, similar to Eq.(\ref{eq:theory_ac}). From this the noise at zero frequency can be computed  by averaging over the time $t$ and integrating over $\tau$.  While there is clearly a link between $W_e$ and $W$, a theoretical effort is called for to make the mathematically sound connection between them.

\emph{Conclusion.} %
We have experimentally demonstrated how a time-dependent voltage acts on the correlations in current fluctuations taken at different times, via the phase accumulated by electrons. This is the corner stone for the generation of quantum microwaves by electron quantum transport. While squeezing and entanglement between electromagnetic fields at two different frequencies has been demonstrated using ac-excited, non-superconducting quantum conductors\cite{Gasse2013,Forgues_Bell,Bartolomei2023}, our time-domain experiment may yield to the detection of similar properties where the quadratures of the electromagnetic field, given by sine and cosine at a given frequency, need to be extended to time-domain \cite{Virally2019} and the excitation voltage engineered accordingly, using Eq.(\ref{Eq_main}).   

\begin{acknowledgments}
We are very grateful to Gabriel Laliberté for his technical help and Eva Dupont-Ferrier for lending us her microwave source. This work was supported by the Canada Research Chair program, the NSERC, the Canada First Research Excellence Fund, the FRQNT, and the Canada Foundation for Innovation.
\end{acknowledgments}

\bibliography{biblio}

\end{document}